\documentclass[aps,twocolumn,tightenlines,english,preprintnumbers,amsmath,amssymb,nofootinbib,superscriptaddress]{revtex4-1}

\usepackage{graphics,bm}
\usepackage{epsfig}
\usepackage{graphicx}
\usepackage{amsmath}
\usepackage{wrapfig}
\usepackage{color}

\newcommand{\beq}{\begin{equation}}
\newcommand{\eeq}{\end{equation}}
\newcommand{\bqa}{\begin{eqnarray}}
\newcommand{\eqa}{\end{eqnarray}}

\def\sumint{\hbox{$\sum$}\!\!\!\!\!\!\int}
\def\square{\vcenter{\vbox{\hrule height.4pt
          \hbox{\vrule width.4pt height4pt
          \kern4pt\vrule width.3pt}\hrule height.4pt}}}

\begin{document}

\title{Consistent regularization and renormalization in models with
inhomogeneous phases}

\author{Prabal Adhikari}
\email{adhika1@stolaf.edu}
\affiliation{St. Olaf College, Physics Department, 1520 St. Olaf Avenue,
Northfield, MN 55057, USA}
\author{Jens O. Andersen}
\email{andersen@tf.phys.ntnu.no}
\affiliation{Department of Physics, Faculty of Natural Sciences,NTNU, 
Norwegian University of Science and Technology, H{\o}gskoleringen 5,
N-7491 Trondheim, Norway}
\affiliation{Niels Bohr International Academy, 
Blegdamsvej 17, Copenhagen 2100, Denmark}
\date{\today}

\begin{abstract}
In many models in condensed matter and high-energy physics,
one finds inhomogeneous phases at high density and low temperature.
These phases are characterized by a spatially dependent 
condensate or order parameter. A proper calculation requires that one
takes the vacuum fluctuations of the model into account.
These fluctuations are ultraviolet divergent and must be regularized.
We discuss different ways of consistently
regularizing and renormalizing 
quantum fluctuations, focusing on momentum cutoff,
symmetric energy cutoff, and 
dimensional regularization. 
We apply these techniques  
calculating the vacuum energy in the NJL model in
1+1 dimensions in the large-$N_c$
limit and in the 3+1 dimensional 
quark-meson model in the mean-field approximation both
for a one-dimensional chiral-density wave.

\end{abstract}

\maketitle

\section{Introduction}
There are many systems in condensed matter and high-energy physics
where some of the phases are inhomogeneous. These are phases where
an order parameter or a condensate 
depends on position.
The simplest case is where only the phase of the order parameter
is varying; the general case is where both magnitude and
phase are functions of position.
The idea of inhomogeneous phases is rather old going back to the 
work of  Fulde and Ferrell as well as by
Larkin and Ovchinikov in the context of superconductors
\cite{loff1,loff2}, density
waves in nuclear matter by Overhauser \cite{over}, and pion condensation by 
Migdal \cite{migdal}. 
In recent years, inhomogeneous phases have been studied
in for example
cold atomic gases \cite{baym}, 
color superconducting phases \cite{incolor,anglo,heman}, 
quarkyonic phases \cite{robd,robd2}, 
as well as chiral condensates 
\cite{sadzi,nakano,nickel,nick2,bubsc,balli,abuki,braun,friman,dirk0,carigo3,dirk}, 
see Refs. \cite{suprev,buballarev}
for recent reviews.
\\ \indent
In the case of QCD, these inhomogeneous phases exist at large
baryon chemical potential $\mu_B$ or isospin 
chemical potential $\mu_I$, and low temperature. In the case of large 
$\mu_B$, they can not be
studied by lattice simulations due to the infamous sign problem and one must
use low-energy models of QCD. Examples of such models are the
Nambu-Jona-Lasinio (NJL) model and the quark-meson (QM) model
or their Polyakov-loop extended versions, the PNJL and PQM models.
Most of the calculations in 3+1 dimensions 
have been inspired by
corresponding calculations in 1+1 dimensions using
ans\"atze 
for the inhomogeneities that are one-dimensional
\cite{full2,fullgn,chiral11,1d1,1d0,symren1,symren2,pionin, thieslast}.
Interestingly, some of these models in 1+1 dimensions can be 
solved exactly in the large-$N_c$ limit and show a rich phase 
diagram \cite{1d1,1d0}.
\\ \indent
When calculating the thermodynamic potential in these models, one faces
ultraviolet divergences due to vacuum fluctuations.
The ultraviolet divergences in the NJL model 
are typically regularized using a sharp momentum cutoff $\Lambda$ \cite{klev}.
However, in the case of inhomogeneous condensates, a naive application
of a momentum cutoff leads
to an incorrect expression for the vacuum energy
in the limit
where the magnitude of the order parameter
vanishes \cite{symren1,symren2}.
Instead, proper time regularization \cite{nakano},
``symmetric energy cutoff regularization'' 
\cite{symren1,symren2} and Pauli-Villars 
regularization have been applied \cite{nickel,nick2,balli,bubsc}.
In this paper, we
will discuss how to use 
momentum cutoff, symmetric energy cutoff, and dimensional
regularization in the case of inhomogeneous phases;
In order to obtain a meaningful expression for the vacuum energy,
it is necessary to perform a unitary transformation 
(that depends on the wave vector)
on the free Hamiltonian and 
subtract the vacuum energy of the noninteracting system.
\\ \indent
In the NJL model, one cannot throw away the quantum fluctuations since
chiral symmetry breaking is induced by them, i.e.
there is no symmetry breaking at tree 
level.\footnote{In the NJL model, there is symmetry breaking for
a coupling constant larger than a critical one, $G_c$, which depends on the
cutoff $\Lambda$.}
This is in contrast to 
the quark-meson (QM) model, where the Higgs mechanism is
implemented by chosing a negative mass term in the tree-level potential. 
In many finite-temperature applications of the quark-meson model,
one ignores the vacuum fluctuations hoping that their effects
on the chiral transition are negligible \cite{nickel,scavenius}.
However, it turns out that vacuum fluctuations are 
important \cite{vac}.
In the two-flavor QM model, the chiral transition
is first order (in the chiral limit) in the whole $\mu_B$--$T$ plane 
without vacuum fluctuations. Adding the quantum fluctuations,
the transition changes from being first order
to being second order at zero baryon chemical potential $\mu_B$, while
it remains first order at zero temperature. Thus, including 
the vacuum fluctuations, the first-order line that starts at $T=0$
ends at a tricritical point somewhere
in the $\mu_B$--$T$ plane. Similarly, when allowing for an inhomogeneous
phase such as a chiral-density wave, it exists in the entire $\mu_B$--$T$ plane 
in the absence of
quantum fluctuations. Including quantum fluctuations,
the inhomogeneous phase emerges from a tricritical point  
and exists in a region of low temperatures down to $T=0$~\cite{bubsc}.
\\ \indent
Conventionally, dimensional regularization has been used in the context
of the quark-meson model.
However, apriori, there is nothing that prevents us from treating the
quark-meson model as a cutoff field theory \cite{lepage}
and regularizing it using a sharp ultraviolet momentum
cutoff $\Lambda$.
Having introduced an ultraviolet cutoff $\Lambda$, one can 
renormalize, i.e. redefine the parameters of the model and take
the limit $\Lambda\rightarrow\infty$ at the end.
In this case, one trades the ultraviolet cutoff for a renormalization
scale $\mu$. In fact, this procedure yields results that are
reminiscent of dimensional regularization in which 
power divergences are set to zero and logarithmic divergences show
up as poles. The poles are then removed by renormalization of the
parameters of the theory.
Either way, there is an ambiguity, since there is a 
dependence on ultraviolet cutoff
$\Lambda$ or the renormalization scale $\mu$.
\\ \indent
The article is organized as follows. In Sec. II. we discuss the
problem of a simple momentum cutoff in the context of an NJL type model in 
1+1 dimensions. We will show that by subtracting the vacuum energy
of the free theory after a unitary transformation,
one can obtain a meaningful vacuum energy using a momentum cutoff, 
an energy cutoff, or dimensional regularization. 
In Sec. III, we show how
to apply these techniques to the quark-meson model in three dimensions.
In Sec. IV, we summarize and discuss our results.

\section{NJL model in 1+1 dimensions}
\subsection{Lagrangian and thermodynamic potential}
The Lagrangian of the NJL model in 1+1 dimensions is
\bqa\nonumber
{\cal L}&=&
\bar{\psi}\left[
i/\!\!\!\partial-m_0
+(\mu+\mbox{$1\over2$}\tau_3\mu_I
)\gamma^{0}
\right]
\psi
\\
&&+{G\over N_c}\left[
(\bar{\psi}\psi)^2+(\bar{\psi}i\gamma^5{\boldsymbol\tau}\psi)^2
\right]\;,
\label{lag1}
\eqa
where $N_c$ is the number of colors 
and $m_0$ is the current quark mass. 
Moreover $\psi$ is 
a color $N_c$-plet, a two-component Dirac spinor as well as a flavor doublet 
\bqa
\psi&=&
\left(
\begin{array}{c}
u\\
d
\end{array}\right)\;.
\eqa
Here
$\mu_B=3\mu=\mbox{$3\over2$}(\mu_u+\mu_d)$ 
and $\mu_I=(\mu_u-\mu_d)$ 
are the baryon and 
isospin chemical potentials
expressed in terms of 
the quark chemical potentials $\mu_u$ and $\mu_d$.
The $\gamma$-matrices are $\gamma^0=\sigma_2$, $\gamma^1=i\sigma_1$, and
$\gamma^5=\gamma^0\gamma^1=\sigma_3$, where ${\sigma}_i$ are the
three Pauli matrices, and $\tau_a$ are the three Pauli 
matrices in 
flavor space. The Lagrangian (\ref{lag1}) has a global $SU(N_c)$
symmetry and for $m_0=\mu_I=0$, it also invariant under
$U_B(1)\times SU_L(2)\times SU_R(2)$.
For $m_0\neq0$ and $\mu_I=0$, the $SU_L(2)\times SU_R(2)$ 
symmetry
is reduced to 
$SU_I(2)$. For $m_0=0$ and $\mu_I\neq0$, 
the symmetry $SU_L(2)\times SU_R(2)$ 
is reduced to $U_{I_3L}(1)\times U_{I_3R}(1)$,
where $I_3$ is the third component of isospin.
If $m_0\neq0$ and $\mu_I\neq0$ 
the $SU_L(2)\times SU_R(2)$ 
symmetry is reduced to $U_{I_3}(1)$.

We next introduce the collective sigma and pion fields
\bqa
\sigma&=&-2{G\over N_c}\bar{\psi}\psi\;,
\\
\pi_a&=&-2{G\over N_c}\bar{\psi}i\gamma^5\tau_a\psi\;.
\eqa
The Lagrangian (\ref{lag1}) then becomes
\bqa\nonumber
{\cal L}&=&
\bar{\psi}\left[
i/\!\!\!\partial
-m_0
+(\mu
+\mbox{$1\over2$}\tau_3\mu_I
)\gamma^{0}-\sigma
-i\gamma^{5}\pi_a\tau_a
\right]\psi
\\ &&
-{N_c(\sigma^2+\pi_a^2)\over4G}\;.
\label{last1}
\eqa
The chiral condensate that we choose is a chiral-density wave
of the form 
\footnote{With a nonzero isospin chemical potential, there is also the
possibilty of a pion condensate $\Delta$. 
For simplicity, we do not include this in the present analysis.}
\bqa
\langle\sigma\rangle&=&M \cos(2bz)-m_0
\label{ansa}
\;,
\\
\langle\pi_3\rangle&=&M\sin(2bz)\;,
\label{back22}
\eqa
where $b$ is a wavevector. 

The Mermin-Wagner-coleman theorem normally forbids spontaneous symmetry
breaking in 1+1 dimensions; however it does not apply in the large $N_c$ limit~\cite{mermin,cole}.
We denote the last term in Eq. (\ref{last1}) by $-V_0$
such that $V_0$ is the tree-level potential.
Note that for nonzero $z$, the crossterm $-{N_cm_0M\cos(2bz)\over2G}$
averages to zero when the spatial extent $L$ of the system
large enough.
This term can then be written as $-{N_cm_0M\delta_{b,0}\over2G}$ 
and the expression for $V_0$ is 
\bqa
V_0&=&{N_c(M^2+m_0^2-2Mm_0\delta_{b,0})\over4G}\;.
\eqa
In the homogeneous case, $V_0={N_c(M-m_0)^2\over4G}$.

With the ansatz (\ref{ansa})--(\ref{back22}), the
Dirac operator $D$ can be written as 
\bqa
D&=&\bar{\psi}\left[
i/\!\!\!\partial
+(\mu+
\mbox{$1\over2$}\tau_3\mu_I)\gamma^0
-Me^{2i\gamma^5\tau_3bz}\right]\psi\;.
\label{dirac}
\eqa
We next redefine the quark fields, $\psi\rightarrow e^{-i\gamma^5\tau_3bz}\psi$
and $\bar{\psi}\rightarrow\bar{\psi}e^{-i\gamma^5\tau_3bz}$.
The Dirac operator then reads
\bqa
D&=&
\left[i/\!\!\!\partial
+(\mu+b^{\prime}\tau_3)\gamma^0-M
\right]\;,
\label{dirac2}
\eqa
where $b^{\prime}=(b+\mbox{$1\over2$}\mu_I)$
and $2b^{\prime}$
is an effective isospin chemical potential.
The transformation of the field $\psi$ amounts to a unitary transformation
of the Dirac Hamiltonian, 
${\cal H}\rightarrow{\cal H}^{\prime}
=e^{i\gamma^5\tau_3bz}{\cal H}e^{-i\gamma^5\tau_3bz}$.
It turns out that there is a spurious dependence on $b$
in the free energy: For some regulators,
the free energy depends on $b$ in the limit $M\rightarrow0$.
However, physical quantities cannot depend on the wavevector 
when the modulus of the condensate
is zero.
This unphysical behavior
of the free energy requires the introduction of a subtraction term
that by construction guarantees that the free energy is independent of
$b$ in the limit $M\rightarrow0$. We will return to this issue below.
There is an additional complication for nonzero isospin chemical
potential since the spurious dependence of $b$ translates into
additional dependence on the isospin chemical potential in the 
free energy. We therefore set $\mu_I=0$ for now and return
to the case of nonzero $\mu_I$ at the end of this section.

Going to momentum space, Eq. (\ref{dirac2}) can be written as
\bqa
D&=&\left[/\!\!\!p
+(\mu+b\tau_3)\gamma^0-M
\right]\;.
\eqa
It is now straightforward to
derive the fermionic spectrum in the background (\ref{back22}).
It is given by the zeros of the Dirac determinant 
and reads \cite{symren1}	
\bqa
E_{\pm}&=&\sqrt{(\sqrt{p^2+M^2}\pm b)^2}\;.
\label{specM}
\eqa
We notice that the lower branch, $E_-$, has zero energy
for nonzero momentum, $p=\pm\sqrt{b^2-M^2}$, if $b>M$.
It is this nonmonotonic behavior that allows for inhomogeneous condensates
at finite chemical potential by lowering the energy and at the same time
populating only the lower branch $E_-$.

We can now integrate over the fermions to obtain the
free energy in the mean-field approximation. 
After integrating over $p_0$,
this yields the standard expression
\bqa
V&=&V_0
-N_c\int_{-\infty}^{\infty}
{dp\over2\pi}(E_++E_-)\;.
\eqa
\subsection{Momentum versus energy cutoff}
The starting point is the one-loop correction to the effective potential,
\bqa
V_1&=&-N_c\int_{-\infty}^{\infty}
{dp\over2\pi}(E_++E_-)\;.
\eqa
If we regulate the integral by a simple momentum cutoff $\Lambda$, we
can write
\bqa
V_1&=&-{N_c\over\pi}\int_0^{\Lambda}(E_++E_-)\,dp\;,
\eqa
It will prove useful to change variable to $u=\sqrt{p^2+M^2}$.
The integral then becomes
\bqa\nonumber
V_1&=&-{N_c\over\pi}\int_M^{\sqrt{\Lambda^2+M^2}}
(|u+b|+|u-b|){udu\over\sqrt{u^2-M^2}}\;.
\\ &&
\label{symmy}
\eqa
We next write $V_1=V_++V_-$.
In the limit of large $\Lambda$, we find 
\bqa\nonumber
V_+&=&
-{N_c\over\pi}\int_M^{\sqrt{\Lambda^2+M^2}}|u+b|{udu\over\sqrt{u^2-M^2}}
\\
&=&-{N_c\over2\pi}\left[
\Lambda^2+2b\Lambda
+{1\over2}M^2\left(
\log{4\Lambda^2\over M^2}+1\right)
\right]\;.
\label{vmer}
\eqa
The other contribution $V_-$ is given by
\bqa
V_-&=&-{N_c\over\pi}\int_M^{\sqrt{\Lambda^2+M^2}}|u-b|{udu\over\sqrt{u^2-M^2}}\;.
\eqa
Here we must be careful
distinguishing between the cases $u>b$ and $u<b$.
In the large-$\Lambda$ limit, one finds 
\bqa\nonumber
V_-&=&-{N_c\over2\pi}
\bigg[
\Lambda^2-2b\Lambda
+{1\over2}M^2\left(
\log{4\Lambda^2\over M^2}+1\right)\bigg]
\\
&&+\theta(b-M)f(M,b)
\;,
\label{vmin1}
\eqa
where we have defined the function $f(M,b)$
\begin{widetext}
\bqa
f(M,b)&=&
-{N_c\over\pi}\left[b\sqrt{b^2-M^2}-
M^2\log{b+\sqrt{b^2-M^2}\over M}\right]
\;.
\eqa
\end{widetext}
The one-loop contribution 
to the free energy
is then given by the sum of 
Eqs. (\ref{vmer}) and (\ref{vmin1}). 
After renormalizing the vacuum energy
by removing the term proportional to $\Lambda^2$, 
the effective potential in the mean-field approximation becomes
\bqa\nonumber
V&=&V_0
-{N_cM^2\over2\pi}\bigg[
\log{4\Lambda^2\over M^2}+1\bigg]
+\theta(b-M)f(M,b)\;.
\\ &&
\label{mom}
\eqa
We note that the terms linear in $b$ cancel and that the final
result is an even function of $b$ as it must, cf. Eq. 
(\ref{symmy}). However, the vacuum energy is 
unbounded below due to the term $b\sqrt{b^2-M^2}$ implying that the system
is unstable.
Moreover, in the limit $M\rightarrow0$, the effective potential reduces to
$V=-{N_cb^2\over\pi}$ (for $m_0=0$).
This is clearly unphysical; the effective potential must be independent
of the wavevector $b$ when the magnitude $M$ of the condensate vanishes.
As pointed out in \cite{symren1,symren2}, the problem is that
the cutoff is imposed on the momentum and not the energy.
Using a momentum cutoff $\Lambda$, the effective cutoff 
on the energy is
$\sqrt{\Lambda^2+M^2}\pm b$, which is different for the two branches
for nonzero $b$.
The idea put forward in
\cite{symren1,symren2} is to use a cutoff $\Lambda$
on the energy rather
than the momentum of the partices, i.e. one restricts the
integration by imposing the same cutoff on the two branches, 
\bqa
E_{\pm}<\Lambda\;.
\eqa
This is referred to as symmetric energy cutoff \cite{symren1,symren2}.
This restriction 
can be expressed as an upper limit for the integration variable $u$
and yields $u<\Lambda\mp b$.
The contribution $V_+$
from $E_+$ is in the large-$\Lambda$ limit then becomes
\bqa\nonumber
V_+&=&-{N_c\over\pi}\int_{M}^{\Lambda-b}|u+b|{udu\over\sqrt{u^2-M^2}}
\\&=&
-{N_c\over2\pi}\left[
\Lambda^2-b^2+{1\over2}M^2\left(\log{4\Lambda^2\over M^2}-1\right)
\right]\;.
\label{eplus}
\eqa
The contribution $V_-$ from $E_-$ is
in the large-$\Lambda$ limit given by
\bqa\nonumber
V_-&=&-{N_c\over\pi}\int_{M}^{\Lambda+b}|u-b|{udu\over\sqrt{u^2-M^2}}
\\ \nonumber
&=&
-{N_c\over2\pi}\left[\Lambda^2
-b^2
+{1\over2}M^2\left(\log{4\Lambda^2\over M^2}-1\right)\right]
\\ &&
+\theta(b-M)f(M,b)
\;.
\label{emin}
\eqa
Note that $V_+=V_-$ when $M>b$.
The final result for the vacuum energy is given by 
the sum of the tree-level term, 
Eqs. (\ref{eplus}) and (\ref{emin})
\bqa\nonumber
V&=&V_0
-{N_c\over2\pi}\left[
-2b^2+
M^2\left(\log{4\Lambda^2\over M^2}-1\right)
\right]
\\ &&
+\theta(b-M)f(M,b)
\;,
\label{efin}
\eqa
where we again have added the vacuum counterterm ${N_c\over\pi}\Lambda^2$.
In the limit $M\rightarrow0$, the
$b$-dependent terms in Eq. (\ref{efin}) drop out and 
the effective potential vanishes (again for $m_0=0$).
Thus the symmetric-energy cutoff provides us with a well-defined
effective potential.

Returning to the momentum cutoff, one can of course simply
subtract the term $-{N_cb^2\over\pi}$ from the effecticve potential, 
but this requires some justification.
The idea is to subtract the
vacuum energy of the noninteracting system, i.e. 
that of a free Fermi gas, as a part of the
renormalization prescription \cite{j+t}.
As explained above, 
the redefinition of the 
quark fields immediately after Eq. (\ref{dirac})
corresponds to a unitary transformation of the 
Hamiltonian of the system. We must therefore also perform 
the same transformation on the free Hamiltonian.
The subtraction term is then obtained by making the
substitution $M\rightarrow m_0$ in Eq. (\ref{mom}).
The total effective potential is given by the sum of 
\bqa\nonumber
V&=&V_0
-{N_cM^2
\over2\pi}\left[
\log{4\Lambda^2\over M^2}+1\right]
+\theta(b-M)f(M,b)
\\&&
-\theta(b-m_0)f(m_0,b)\;,
\label{altern2}
\eqa
where we have dropped terms that depend on $\Lambda$
and $m_0$.
Taking the limits $m_0\rightarrow0$ and $M\rightarrow0$ (in this order),
we see that all the $b$-dependent terms cancel as they should.
We note in passing that the subtraction term is not
unique. We could have subtracted the vacuum energy
of a massless Fermi gas, $V_{\rm sub}=-{N_cb^2\over\pi}$, which 
was done in Ref. \cite{symren1}.
If we use this prescription, the expression for the vacuum energy becomes
\bqa\nonumber
V&=&V_0
-{N_c\over2\pi}\left[-2b^2+
M^2\left(\log{4\Lambda^2\over M^2}+1\right)\right]
\\ &&
+\theta(b-M)f(M,b)\;.
\eqa
\subsection{Dimensional regularization}
Let us next consider the vacuum energy using dimensional regularization.
The one-loop energy is given by the sum of the two terms
\bqa
V_{\pm}&=&-N_c\int_pE_{\pm}\;,
\eqa
where the integral is defined in $d=1-2\epsilon$ dimensions:
\bqa
\int_p&=&\left({e^{\gamma_E}\Lambda^2\over4\pi}\right)^{\epsilon}
\int{d^dp\over(2\pi)^d}\;.
\eqa
Here $\Lambda$ is the renormalization scale associated with 
the $\overline{\rm MS}$ renormalization scheme.
We first change variables $u=\sqrt{p^2+M^2}$ and integrate over angles,
This yields
\bqa\nonumber
V_{\pm}&=&-{N_c\left(e^{\gamma_E}\Lambda^2\right)^{\epsilon}
\over\sqrt{\pi}\Gamma({1\over2}-\epsilon)}
\int_{M}^{\infty}|u\pm b|
{u\,du\over(u^2-M^2)^{{1\over2}+\epsilon}}\;.
\\
\label{vppp}
\eqa
We first consider the contribution $V_+$
to the effective potential from $E_+$. We find
\bqa
V_+&=&{N_cM^2\over4\pi}
\left({e^{\gamma_E}\Lambda^2\over M^2}\right)^{\epsilon}
\Gamma(-1+\epsilon)\;.
\label{vplus}
\eqa
The contribution $V_+$ is independent of $b$, which is most easily
understood by going back to momentum space in Eq. (\ref{vppp}).
The $b$-dependence is then given by an integral over $p$
with no mass scale multiplied by
$b$, and this integral vanishes in dimensional regularization.

The contribution $V_-$ from the negative solution $E_-$ is
given by
\begin{widetext}
\bqa\nonumber
V_-&=&-{N_c\left(e^{\gamma_E}\Lambda^2\right)^{\epsilon}
\over\sqrt{\pi}\Gamma({1\over2}-\epsilon)}\int_{M}^{\infty}
{|u-b|u\,du\over(u^2-M^2)^{{1\over2}+\epsilon}}\;.
\\  \nonumber
&=&
-{N_c\left(e^{\gamma_E}\Lambda^2\right)^{\epsilon}\over
\sqrt{\pi}\Gamma({1\over2}-\epsilon)}
\left[
\int_{M}^{\infty}
{(u-b)u\,du\over(u^2-M^2)^{{1\over2}+\epsilon}}
+2\theta(b-M)
\int_{M}^{b}
{(b-u)u\,du\over(u^2-M^2)^{{1\over2}+\epsilon}}
\right]
\\
&=&
{N_cM^2\over4\pi}
\left({e^{\gamma_E}\Lambda^2\over M^2}\right)^{\epsilon}
\Gamma(-1+\epsilon)
+\theta(b-M)f(M,b)\;,
\label{vmin}
\eqa
where the second integral has been evaluated directly
in one dimension.
We note that the first term in Eq. (\ref{vmin}) is equal to $V_+$.
\end{widetext}
The total vacuum energy is given by the sum of 
the tree-level term,  
Eqs. (\ref{vplus}) and (\ref{vmin}). 
Expanding this expression in powers of $\epsilon$, we obtain
\bqa\nonumber
V&=&V_0
-{N_cM^2\over2\pi}
\left({\Lambda^2\over M^2}\right)^{\epsilon}
\bigg[{1\over\epsilon}+1\bigg]
\\ &&
+\theta(b-M)f(M,b)
\;.
\eqa
The pole in $\epsilon$ is removed by renormalizing 
the quark mass $m_0$ and
the constant coupling $G$. 
This is carried out by the substitutions
$m_0\rightarrow Z_{m_0}m_0$, 
and ${1\over G}\rightarrow Z_{G^{-1}}{1\over G}$
where the mass and inverse coupling 
renormalization constants are
\bqa
\label{ren1}
Z_{m_0}&=&\left[1+{2G\over\pi\epsilon}\right]^{-1}\;,
\\
Z_{G^{-1}}&=&\left[1+{2G\over\pi\epsilon}\right]\;.
\label{ren2}
\eqa
Note that $Z_{G^{-1}}={Z_G}^{-1}$ and that
the ratio $m_0\over G$ is the same for bare and renormalized 
quantites since $Z_{m_0}Z_{G^{-1}}=1$.
This yields the renormalized effective potential in the
mean-field approximation
\bqa\nonumber
V&=&V_0
-{N_cM^2\over2\pi}\bigg[\log{\Lambda^2\over M^2}+1\bigg]
+\theta(b-M)f(M,b)\;.
\\&&
\label{prop}
\eqa
We note in passing that the substitutions (\ref{ren1})--(\ref{ren2})
correspond to 
a nonperturbative renormalization. In perturbation theory, it 
amounts
to summing an infinite series of diagrams from all orders of 
perturbation theory. This can for example 
be seen by analyzing the model in terms of the two-particle
irreducible effective action formalism to leading order in
the $1/N_c$-expansion in analogy with the bosonic case in three dimensions
\cite{gert,jens}. Moreover, the running
coupling $G$ and the running mass $m_0$
satisfy the renormalization group equations
\bqa
\label{rg1}
\Lambda{dG\over d\Lambda}&=&-{4G^2\over\pi}\;,
\\
\Lambda{dm_0\over d\Lambda}&=&-{4m_0G\over\pi}\;.
\label{rg2}
\eqa

However, Eq. (\ref{prop}) is still problematic. Taking the
limit $M\rightarrow0$, we find $V=-{N_cb^2\over\pi}$ (for $m_0=0$)
which is 
unphysical. In order to understand the source of the problem, we must
go back to the contribution $V_{\pm}$ and take the limit $M\rightarrow0$:
\bqa
V_{\pm}&=&-{N_c\over\pi}\left (\frac{e^{\gamma_{E}}\Lambda^{2}}{4\pi}\right )^{\epsilon}\int_0^{\infty}|p\pm b|p^{-2\epsilon}\,dp\;.
\eqa
$V_+$ can be written as a sum of two integrals in which there is no mass
scale. These integrals are then set to zero in dimensional regularization.
In $V_-$, the wavevector $b$ is the only scale in the integral and 
according to the
rules of dimensional regularization, the integral will be proportional
to the appropriate power of 
$b$.\footnote{Alternatively, we note that 
$V_-$ is propertional to $\int_0^{\infty}(p-b)p^{-2\epsilon}\,dp 
+2\int_0^b(b-p)p^{-2\epsilon}dp$, 
where only the latter integral is nonzero.} 
Dimensional analysis gives
$V_-\sim b^{2-2\epsilon}$. The coefficient is finite and in 
the limit $\epsilon\rightarrow0$ one finds $V_-=-{N_cb^2\over\pi}$.
This expression is exactly the vacuum energy of $N_c$ massless 
fermions after a unitary transformation of the Hamiltonian.
As in case of a momentum cutoff, we subtract the vacuum energy in the
normal phase as part of renormalization procedure. This is given by
the second and third term in Eq. (\ref{prop}) after the substitution
$M\rightarrow m_0$. Dropping trivial
terms that depend on $m_0$ and $\Lambda$, we find
\bqa\nonumber
V&=&V_0
-{N_cM^2
\over2\pi}\left[
\log{\Lambda^2\over M^2}+1\right]
+\theta(b-M)f(M,b)
\\&&
-\theta(b-m_0)f(m_0,b)\;.
\label{altern4}
\eqa
Our results for the vacuum energy
in the three regularization regularization schemes are given by 
Eqs. (\ref{efin}), (\ref{altern2}), and (\ref{altern4}).
In the case of nonzero isospin chemical potential,
these results are still somewhat problematic. Consider the free energy
Eq. (\ref{efin}) in the case $b=0$. It does not reduce to the
free energy of a massive Fermi gas at $T=0$ due to the extra
term $-{N_c\over4\pi}\mu_I^2$.
The problem can be solved simply by adding the $b$-independent term
${N_c\over4\pi}\mu_I^2$  to the vacuum energy \cite{symren1}.
This term or $\theta(\mbox{$1\over2$}\mu_I-m_0)f(m_0,\mbox{$1\over2$}\mu_I)$
should be added to the vacuum energy calculated in the
momentum cutoff scheme (\ref{altern2})
and dimensional regularization (\ref{altern4}).

We close this section by noting that in the chiral limit,
the gap equation ${dV\over dM}=0$, 
in the vacuum where $b=0$, has two solutions, either $M_0=0$ or
\bqa
M_0&=&\Lambda e^{-{\pi\over4G}}\;.
\label{nona}
\eqa
Using the renormalization group equation (\ref{rg1}), it is easy
to verify that  $M_0$ is renormalization group invariant.
The nonanalytic behavior of $M_0$ as a function of $G$ shows that it is
a nonperturbative result. As mentioned above, it corresponds to 
summing an infinite series of diagrams from all order of perturbation theory.

Eq. (\ref{nona}) can be used to trade the cutoff or the renormalization 
scale for the mass scale $M_0$. In dimensional regularization, we find
\bqa
V&=&-{N_cM^2\over2\pi}\left[\log{M_0^2\over M^2}+1\right].
\label{rr}
\eqa
The RG-invariance of $M_0$ implies the RG-invariance of
$V$ in (\ref{rr}).
As pointed out in \cite{symren1},
the unrenormalized expression for the vacuum energy contains a 
dimensionless parameter $G$, while the renormalized result (\ref{rr})
contains a dimensionful mass scale $M_0$.
This is an example of dimensional transmutation.
\section{Quark-meson model}
\subsection{Lagrangian and thermodynamic potential}
The Euclidean Lagrangian of the two-flavor quark-meson model is 
\bqa\nonumber
{\cal L}&=&
{1\over2}\left[(\partial_{\mu}\sigma)^2
+(\partial_{\mu}{\boldsymbol \pi})^2\right]
+{1\over2}m^2(\sigma^2+{\boldsymbol\pi}^2)
\\&&
\nonumber
-h\sigma
+{\lambda\over24}(\sigma^2+{\boldsymbol\pi}^2)^2
\\ && \nonumber
+\bar{\psi}
\left[
/\!\!\!\partial
-(\mu+\mbox{$1\over2$}\tau_3\mu_I)\gamma^0
+g(\sigma+i\gamma^5{\boldsymbol\tau}\cdot{\boldsymbol\pi})\right]\psi\;,
\\ 
\label{lag}
\eqa
where $\psi$ is 
a color $N_c$-plet, a four-component Dirac spinor as well as a flavor doublet 
\bqa
\psi&=&
\left(
\begin{array}{c}
u\\
d
\end{array}\right)\;.
\eqa
Here $\mu_B=3\mu=\mbox{$3\over2$}(\mu_u+\mu_d)$ 
and $\mu_I=(\mu_u-\mu_d)$ 
are the baryon and 
isospin chemical potentials expressed in terms of 
the quark chemical potentials $\mu_u$ and $\mu_d$,
$\tau_i$ ($i=1,2,3$) are the Pauli matrices in flavor space, and 
${\boldsymbol\pi}=(\pi_1,\pi_2,\pi_3)$.

Apart from the global $SU(N_c)$ symmetry, 
the Lagrangian~(\ref{lag}) 
has a 
$U(1)_B\times SU(2)_L\times SU(2)_R$ symmetry for 
$h=0$ and a $U(1)_B\times SU(2)_V$ symmetry
for $h\neq0$. 
When $\mu_u\neq\mu_d$, this symmetry is reduced to 
$U(1)_B\times U_{I_3L}(1)\times U_{I_3R}(1)$ for $h=0$ and
$U(1)_B\times U_{I_3}(1)$ for $h\neq0$. In the remainder of this
paper we take $h=0$, i.e. we work in the chiral limit.
We also set $\mu_u=\mu_d$.

In order to study inhomogeneous phases, we must make an ansatz for
the space-time dependence of the mesonic mean fields.
In the literature, mainly one-dimensional modulations have been considered,
for example chiral-density waves (CDW) and soliton lattices.
Since the results seem fairly independent of the 
modulation \cite{bubsc}, we opt for the simplest, namely 
a chiral-density wave. The ansatz is
\bqa
\sigma(z)=
\phi_0\cos(qz)
\;,
\hspace{0.5cm}
\pi_3(z)=
\phi_0\sin(qz)\;,
\label{back}
\eqa
where $\phi_{0}$ is the magnitude of the condensate and 
$q$ is the wavevector. The mean fields can be
be combined into a complex order parameter 
as $M(z)=g[\sigma(z)+i\pi_3(z)]$ or
\bqa
M(z)&=&\Delta e^{iqz}\;,
\label{mz}
\eqa
where we have introduced $\Delta=g\phi_0$.
After averaging over a sufficiently large volume $V_3$ in three dimensions,
the tree-level effective potential is then
\bqa
V_0&=&
{1\over2}q^2{\Delta^2\over g^2}
+{1\over2}m^2{\Delta^2\over g^2}
+{\lambda\over24}{\Delta^4\over g^4}\;.
\eqa
In analogy with the previous example, we can derive the spectrum by finding
the zeros of the Dirac determinant. The result is~\cite{dautry}
\bqa
E_{\pm}&=&\sqrt{
\left(\sqrt{p_{\parallel}^2+\Delta^2}\pm{q\over2}\right)^2+p_{\perp}^2
}\;,
\label{disp}
\eqa
where $p_{\perp}^2=p_1^2+p_2^2$, and $p_{\parallel}=p_3$.
Note that the lower branch has a vanishing minimum,
$E_-=0$, for nonzero momentum
$p_{\parallel}=\pm\sqrt{{q^2\over4}-\Delta^2}$
and $p_{\perp}=0$ in the case ${q\over2}>\Delta$.
It is this nonmonotonic behavior that allows for inhomogeneous condensates at
finite density; it may be energetically favorable
for the system to develop a nonzero value of $q$ and populate only the
lower branch $E_-$.
Although inhomogeneous phases are possible only for nonzero chemical
potentials, the vacuum energy
is independent of $\mu_f$ so the chemical potentials play no role
in the calculation below.

\subsection{Energy and momentum cutoff}
The vacuum part of the
one-loop contribution to the effective potential is given by
the expression
\bqa
V_1&=&-2N_c\int_p(E_++E_-)\;,
\label{v3d}
\eqa
where the integral is in three spatial dimensions.
We first use an energy cutoff 
to evaluate Eq. (\ref{v3d}).
In the case of $E_-$, we must 
distinguish between the cases $\sqrt{p_{\parallel}^2+\Delta^2}-{q\over2}>0$, 
and $\sqrt{p_{\parallel}^2+\Delta^2}-{q\over2}<0$.
As in 1+1 dimensions, there
is an extra term in the case 
$\sqrt{p_{\parallel}^2+\Delta^2}-{q\over2}<0$, which we
denote by $f(\Delta,q)$.
We first integrate over $p_{\perp}$ from zero to 
$p_{\perp}^{\rm max}=\sqrt{\Lambda^2-(u\pm {q\over2})^2}$, and then integrate
over $u$
($u=\sqrt{p_{\parallel}^2+\Delta^2}$)
from $u=\Delta$ to $u=\Lambda\mp {q\over2}$
(upper sign for $E_+$ and lower sign for $E_-$).
The expressions for the integrals are
\begin{widetext}
\bqa\nonumber
V_+&=&
-2N_c\int_pE_+
\\ \nonumber
&=&
-{2N_c\over(4\pi)^2}
\left[
{1\over6}\sqrt{\left(\Lambda-{q\over2}\right)^2-\Delta^2}
\left[
\left(\Lambda-{q\over2}\right)(12\Lambda^2+4\Lambda q+q^2)-
\Delta^2(6\Lambda+13q)\right]\right.\\ 
&&\left.
-(\Delta^4+\Delta^2q^2)
\log{(\Lambda-{q\over2}+\sqrt{(\Lambda-{q\over2})^2-\Delta^2})
\over\Delta}
\right]\;,
\label{ip}
\\ \nonumber
V_-&=&-2N_c\int_pE_-\\&=&
\nonumber
-{2N_c\over(4\pi)^2}\left[
{1\over6}\sqrt{\left(\Lambda+{q\over2}\right)^2-\Delta^2}\left[
\left(\Lambda+{q\over2}\right)(12\Lambda^2-4\Lambda q+q^2)-
\Delta^2(6\Lambda-13q)\right]
\right. \\ &&\left. 
-(\Delta^4+\Delta^2q^2)\log{(\Lambda+{q\over2}+\sqrt{(\Lambda+{q\over2})^2
-\Delta^2})\over\Delta}
\right]
+\theta(\mbox{$q\over2$}-\Delta)f(\Delta,q)\;,
\label{im}
\eqa
where the function $f(\Delta,q)$ is defined as
\bqa
f(\Delta,q)&=&
{N_c\over3(4\pi)^2}
\left[q\sqrt{{q^2\over4}-\Delta^2}(26\Delta^2+q^2)
-12\Delta^2(\Delta^2+q^2)\log{q+2\sqrt{{q^2\over4}-\Delta^2}\over2\Delta}
\right]
\;.
\label{fdef}
\eqa
In the limit $\Delta\rightarrow0$, 
$V_++V_-$ reduces to $-{8N_c\over(4\pi)^2}\Lambda^4$
showing that the thermodynamic potential is
independent of $q$ in this limit.
Subtracting this term 
then corresponds to a trivial renormalization of the vacuum energy.
In the limit $\Lambda\rightarrow\infty$, the sum of
(\ref{ip}) and (\ref{im}) behaves as
\bqa
V_++V_-
&=&-{2N_c\over(4\pi)^2}\left[
-4\Lambda^2\Delta^2
-q^2\Delta^2\left[\log{4\Lambda^2\over\Delta^2}-2\right]
-\Delta^4\left[\log{4\Lambda^2\over\Delta^2}-{1\over2}\right]
+{1\over12}q^4\right]+f(\Delta,q)\;,
\label{divdiv}
\eqa
in agreement with the result 
first obtained by Broniowski and Kutschera \cite{kut}.

%
Let us briefly discuss the calculation of the vacuum energy using a
momentum cutoff $\Lambda$.   
Integrating Eqs. (\ref{ip}) and (\ref{im})
and taking the limit $\Delta\rightarrow0$, we find
$V_++V_-=-{8N_c\over(4\pi)^2}(\Lambda^4+{2\over3}q^2\Lambda^2+{1\over15}q^4)$, 
which
must be subtracted. For large $\Lambda$, the final result is
\bqa
V_++V_-
&=&-{2N_c\over(4\pi)^2}\bigg\{
4\Lambda^2\Delta^2
-\Delta^2q^2\left[\log{4\Lambda^2\over\Delta^2}-{5\over3}\right]
-\Delta^4\left[\log{4\Lambda^2\over\Delta^2}-{1\over2}\right]
+{1\over12}q^4\bigg\}
+f(\Delta,q)
\;.
\label{cutiff}
\eqa

\end{widetext}
Comparing Eqs. (\ref{divdiv}) and (\ref{cutiff}), we
see that the coefficients of some of the terms are different.
However, the
coefficients of the logarithmic terms are identical.


\subsection{Dimensional regularization}
We next consider dimensional regularization. The integrals needed are
\bqa
V_{\pm}&=&-2N_c\int_pE_{\pm}\;,
\eqa
where the integral is in $d=3-2\epsilon$ dimensions,
\\
\bqa\nonumber
\int_p&=&\left({e^{\gamma_E}\Lambda^2\over4\pi}\right)^{\epsilon}
\int{d^dp\over(2\pi)^d}
\\ &=&
\left({e^{\gamma_E}\Lambda^2\over4\pi}\right)^{\epsilon}
\int_{p_{\perp}}{d^{d-1}p_{\perp}\over(2\pi)^{d-1}}
\int_{p_{\parallel}}{dp_{\parallel}\over2\pi}
\;.
\eqa
\\
We first integrate over angles in the $(p_1,p_2)$-plane 
and introduce the variable
$u=\sqrt{p_{\parallel}^2+\Delta^2}$.
The integral then becomes
\bqa\nonumber
V_{\pm}&=&-{N_c(e^{\gamma_E}\Lambda^2)^{\epsilon}\over\pi^{2}\Gamma(1-\epsilon)}
\int_{\Delta}^{\infty}
{u\,du\over\sqrt{u^2-\Delta^2}}
\\ &&
\times
\int_{0}^{\infty}dp_{\perp}
\sqrt{\left(u\pm{q\over2}\right)^2+p_{\perp}^2}\,p_{\perp}^{1-2\epsilon}
\;.
\label{integralane}
\eqa
In contrast to calculation in the 1+1 dimensional NJL model, we were
not able to calculate directly in dimensional regularization
the vacuum energy given by $V_1=V_++V_-$. We therefore use another strategy.
In order to isolate the ultraviolet divergences, we 
expand the integrand in powers of $q$ and identify appropriate
subtraction terms. This yields
\begin{widetext}
\bqa
\sqrt{\left(u\pm{q\over2}\right)^2+p_{\perp}^2}&=&
\sqrt{u^2+p^2_{\perp}}\pm{uq\over2\sqrt{u^2+p_{\perp}^2}}
+{q^2p_{\perp}^2\over8(u^2+p_{\perp}^2)^{3\over2}}
\mp{q^3p_{\perp}^2u\over16(u^2+p_{\perp}^2)^{5\over2}}
+{q^4p_{\perp}^2(4u^2-p_{\perp}^2)\over128(u^2+p_{\perp}^2)^{7\over2}}+...
\label{sub22}
\eqa
We denote the right-hand side of (\ref{sub22}) by ${\rm sub}_{\pm}(u,p_{\perp})$
and write the integrals in (\ref{integralane}) as
\bqa
V_{\pm}&=&V_{\rm div\pm}+V_{\rm fin\pm}-V_{\rm fin\pm}\;,
\eqa
where 
\bqa
V_{\rm div\pm}&=&
-{N_c(e^{\gamma_E}\Lambda^2)^{\epsilon}\over\pi^{2}\Gamma(1-\epsilon)}
\int_{\Delta}^{\infty}
{u\,du\over\sqrt{u^2-\Delta^2}}
\int_{0}^{\infty}
{\rm sub}_{\pm}(u,p_{\perp})p_{\perp}^{1-2\epsilon}\,dp_{\perp}
\;,
\\
V_{\rm fin\pm}&=&
-{N_c(e^{\gamma_E}\Lambda^2)^{\epsilon}\over\pi^{2}\Gamma(1-\epsilon)}
\int_{\Delta}^{\infty}
{u\,du\over\sqrt{u^2-\Delta^2}}
\int_{0}^{\infty}
\left[\sqrt{\left(u\pm{q\over2}\right)^2+p_{\perp}^2}
-{\rm sub}_{\pm}(u,p_{\perp})\right]
p_{\perp}^{1-2\epsilon}\,dp_{\perp}\;.
\eqa
The integral $V_{\rm fin\pm}$ can now be calculated directly in three dimensions.
After integrating over $p_{\perp}$, we find
\bqa
V_{\rm fin\pm}&=&
-{N_c
\over3\pi^{2}}
\int_{\Delta}^{\infty}
{u\,du\over\sqrt{u^2-\Delta^2}}(u\pm\mbox{$q\over2$})^2\left[
\left(u\pm\mbox{$q\over2$}\right)-\big|u\pm\mbox{$q\over{2}$}\big|
\right]\;.
\eqa
Thus $V_{\rm fin+}$ vanishes identically
and $V_{\rm fin-}$ becomes
\bqa \nonumber
V_{\rm fin -}
&=&
-{2N_c\over3\pi^2}\int_{\Delta}^{\infty}
{u\,du\over\sqrt{u^2-\Delta^2}}(u-\mbox{$q\over2$})^3
\theta(\mbox{$q\over2$}-\Delta)
\\ \nonumber
&=&{N_c\over3(4\pi)^2}
\left[q\sqrt{{q^2\over4}-\Delta^2}(26\Delta^2+q^2)
-12\Delta^2(\Delta^2+q^2)\log{q+2\sqrt{{q^2\over4}-\Delta^2}\over2\Delta}
\right]
\theta(\mbox{$q\over2$}-\Delta)
\\ 
&=&
f(\Delta,q)\;.
\label{finf}
\eqa
We next  integrate $V_{\rm div\pm}$ using dimensional regularization.
This is done by first integrating over $p_{\perp}$ and then over $u$.
This yields
\bqa\nonumber
V_{\rm div}&=&V_{\rm div+}+V_{\rm div-}
\\
&=&
{2N_c\over(4\pi)^2}
\left({e^{\gamma_E}\Lambda^2\over\Delta^2}\right)^{\epsilon}
\left[
2\Delta^4
\Gamma(-2+\epsilon)+q^2\Delta^2\Gamma(\epsilon)
+{q^4\over12}(-1+\epsilon)\Gamma(1+\epsilon)
\right]\;.
\eqa
Expanding $V_{\rm div}$ to zeroth order in powers of $\epsilon$, we obtain
\bqa
V_{\rm div}&=&
{2N_c\over(4\pi)^2}\left({\Lambda^2\over\Delta^2}
\right)^{\epsilon}
\left[
\left({1\over\epsilon}+{3\over2}\right)\Delta^4
+{1\over\epsilon}\Delta^2q^2
-{q^4\over12}
+{\cal O}(\epsilon)
\right]\;.
\label{divf}
\eqa
The one-loop effective potential is then given by the sum of Eqs. (\ref{finf})
and (\ref{divf}). It 
contains poles in $\epsilon$, which are
removed by mass and coupling-constant renormalization.
This  amounts to making the substitutions
$m^2\rightarrow Z_{m^2}m^2$, 
$\lambda\rightarrow Z_{\lambda}\lambda$, 
and
$g^2\rightarrow Z_{g^2}g^2$, 
where 
\bqa
Z_{m^2}=
1+{4N_cg^2\over(4\pi)^2\epsilon}\;,
\hspace{1cm}
Z_{\lambda}=1+
{8N_c\over(4\pi)^2\epsilon}
\left[
\lambda g^2-6g^4\right]
\;,
\label{dl}
\hspace{1cm}
Z_{g^2}=1+{4N_cg^2\over(4\pi)^2\epsilon}
\;.
\eqa
After renormalization, the 
effective potential in the mean-field approximation reads
\bqa
V&=&
{1\over2}q^2{\Delta^2\over g^2}
+{1\over2}m^2{\Delta^2\over g^2}
+{\lambda\over24}{\Delta^4\over g^4}
+{2N_cq^2\Delta^2\over(4\pi)^2}\log{\Lambda^2\over\Delta^2}
+{2N_c\Delta^4\over(4\pi)^2}\left[\log{\Lambda^2\over\Delta^2}+{3\over2}
\right]
-{N_cq^4\over6(4\pi)^2}
+f(\Delta,q)\;.
\eqa
\end{widetext}
In contrast to the example in 1+1 dimensions, we need not subtract
a term proportional to the appropriate power of the wavevector
(here $q^4$)
to obtain an effective potential with the right properties. 
The reason is simply that the vacuum energy
is independent of $q$ for $\Delta=0$.

We close this section by discussing how dimensional regularization
can be used in conjunction with a Landau-Ginzburg (GL) analysis of the
quark-meson model. In this case we expand the effective potential
in powers of $\Delta$ and its derivatives. 
Up to a temperature-dependent constant, we find
\bqa\nonumber
V&=&{1\over2}q^2\frac{\Delta^2}{g^{2}}+
{1\over2}m^2\frac{\Delta^2}{g^{2}}+{\lambda\over24}\frac{\Delta^4}{g^{4}}+
\beta_1\Delta^2
\\ &&
+\beta_2\Delta^4+\beta_3(\nabla\Delta)^2+...\;,
\eqa
where the coefficients are
\bqa
\beta_1&=&-4N_c\sumint_{\{P\}}{1\over P^2}\;,
\\
\beta_2&=&2N_c\sumint_{\{P\}}{1\over P^4}\;,
\\ 
\beta_3&=&-N_c\sumint_{\{P\}}\left[
{4p_{\parallel}^2\over P^6}-{3\over P^4}
\right]\;.
\eqa
Here, the sum-integral is defined by
\bqa
\sumint_{\{P\}}&=&
\left({e^{\gamma_E}\Lambda^2\over4\pi}\right)^{\epsilon}
T\sum_{\{P_0\}}\int{d^{d}p\over(2\pi)^d}\;,
\eqa
where $P_0=(2n+1)\pi T+i\mu$ are the fermionic Matsubara frequencies
with $n=0,\pm1,\pm2...\;$.
Using integrating by parts in $d=3-2\epsilon$ dimension, it is straightforward
to show that $\beta_2=\beta_3$.
This result was first obtained in \cite{bubsc} using 
Pauli-Villars regularization. For the special value of the sigma mass
$m_{\sigma}=2\Delta$,
it was shown in \cite{bubsc} that this implies that the tricritical 
point is actually a Lifschitz point. In the NJL model this is always
the case when using a regulator where the total derivative vanishes
\cite{nickel}.
Due to infinite surface terms, such an expansion is problematic
in the case of a momentum 
cutoff. This problem is avoided in the NJL model in 
1+1 dimensions, since the coefficients in the GL functional are finite.

\section{Summary and Discussion}
In this paper, we have for the first time
discussed momentum cutoff regularization, 
symmetric energy cutoff regularization,
and dimensional regularization in the context of one-dimensional
inhomogeneities in the NJL and QM models.
We have shown that all regularization schemes can be used
to define a physically meaningful vacuum energy.
In the case of symmetric energy cutoff regularization, the result
is independent of the wavevector when the magnitude of the 
condensate vanishes, while in the other cases one must subtract
a wavevector-dependent term. We propose to subtract such a term
for all regularizations as a part of the renormalization procedure.
In the examples considered in this paper, an appropriate term is
the Hamiltonian of a free Fermi gas after a unitary transformation.
After this subtraction, one must also add a term
that depends on the isospin chemical potential in order to obtain
the correct expression for the free energy and isospin density in
the limit $b=0$.

We have also briefly discussed finite temperature and a Ginzburg-Landau
analysis of critical points.
Due to the absence of surface terms in the coefficients of the
GL functional, dimensional regularization can always be used 
in the analysis of critical points. The application of momentum cutoff or
symmetric energy cutoff at finite temperature is restricted to the
cases where the GL coefficients are finite, for example
the NJL model in 1+1 dimensions.
Results for the phase diagram of 
the 1+1 dimensional NJL model
is presented in \cite{jensny}.
\\ \indent
There are other regularization schemes that we have briefly mentioned, 
namely Schwinger's proper time regularization and 
Pauli-Villars regularization. The latter method was successfully 
applied to the problem of inhomogeneous phases in the NJL model \cite{nickel}
and the QM model \cite{bubsc}, where the equality 
of the two coefficients $\beta_2$ and $\beta_3$ was shown.
In other words, Pauli-Villars regularization has the same virtues
as dimensional regularization although the final expressions
for renormalized quantities are not so compact.
\\ \indent
It is often argued that since the NJL model in three dimensions
is ``nonrenormalizable", 
one cannot use dimensional regularization but is forced to
use cutoff (momentum or energy) regularization or Pauli-Villars
regularization. We disagree with this view. Nonrenormalizability alone
cannot be an argument against applying dimensional regularization
since it has been applied succesfully to nonrenormalizable models.
For example, it has been used in  
chiral perturbation theory \cite{chpt}
and in the theory of
weakly interacting Bose gases and Bose condensation, both involving
nonrenormalizable field theories \cite{eric,jensbose}.

\section*{Acknowledgments}
The authors would like to thank Tomas Brauner,
Michael Buballa and Bernd-Jochen
Schaefer for valuable discussions. 
The authors
would like to thank the Niels Bohr International Academy for its hospitality
during the latter stages of this work. P.A. would like to thank Professor Sujeev Wickramasekera (Grinnell College) for his mentorship and inspiring scientific work.

\bibliography{refs}{}

\begin{thebibliography}{99}

\bibitem{loff1}
P. Fulde and R. A. Ferrell, Phys. Rev. {\bf 135}, A550, (1964).
\bibitem{loff2}
A. Larkin and Y. Ovchinnikov, Zh. Eksp. Teor. Fiz. {\bf 47},
1136 (1964).

\bibitem{over}
A. W. Overhauser, Phys. Rev. Lett. {\bf 4}, 415 (1960).

\bibitem{migdal}
A. B. Migdal, Rev. Mod. Phys. {\bf 50},  107 (1978).
 
\bibitem{baym}
K. Maeda, T. Hatsuda, and G. Baym, 
Phys. Rev. A {\bf 87},  021604 (2013).

\bibitem{incolor}
M. G. Alford, J. A. Bowers, and K. Rajagopal, 
Phys. Rev. D {\bf 63},  074016 (2001).


\bibitem{anglo}
R. Anglani, G. Nardulli, M. Ruggieri, and M. Mannarelli,
Phys. Rev. D {\bf 74}, 074005 (2006).

\bibitem{heman}
L. He, M. Jin, and P.-F. Zhuang,
Phys. Rev. D {\bf 75}, 036003 (2007).


\bibitem{robd}
T. Kojo, Y. Hidaka, L. McLerran, and R. D. Pisarski,
Nucl. Phys. A {\bf 843}, 37  (2010); ibid {\bf 875}, 94, (2012).

\bibitem{robd2}
T. Kojo, R. D. Pisarski, and A.M. Tsvelik,
Phys. Rev. D {\bf 82}, 074015 (2010).










\bibitem{sadzi}
M. Sadzikowski and W. Broniowski,
Phys. Lett. B {\bf 488}, 63 (2000).



\bibitem{nakano}
E. Nakano and T. Tatsumi, 
Phys. Rev. D {\bf 71}, 114006 (2005).


\bibitem{nickel}D. Nickel,
Phys. Rev. {\bf D} 80, 074025 (2009).

\bibitem{nick2}
D. Nickel,
Phys. Rev. Lett. {\bf 103}, 072301 (2009).

\bibitem{balli}
S. Carignano, D. Nickel, and M. Buballa
Phys. Rev. D {\bf 82}, 054009 (2010).

\bibitem{bubsc}
S. Carignano, M. Buballa, and B.-J. Schaefer,
Phys. Rev.  D {\bf 90}, 014033 (2014).

\bibitem{abuki}
H. Abuki,
Phys. Lett. {\bf B} 728, 427 (2014).

\bibitem{braun}	
J. Braun, S. Finkbeiner, F. Karbstein, and D. Roscher 
Phys. Rev. D {\bf 91},116006 (2015).

\bibitem{friman}
T.-G. Lee, E. Nakano, Y. Tsue, T. Tatsumi, B. Friman,
Phys. Rev. D {\bf 92}, 034024 (2015).

\bibitem{dirk0}
A. Heinz, F. Giacosa, and D. H. Rischke,
Nucl.Phys. A {\bf 933},  34 (2015).

\bibitem{carigo3}
M. Buballa and S. Carignano, 
Eur. Phys. J. A {\bf 52}, 57 (2016).

\bibitem{dirk}
A. Heinz, F. Giacosa, M. Wagner, and D. H. Rischke,
Phys. Rev.  D {\bf 93}, 014007 (2016) 



\bibitem{suprev}
R. Anglani, R. Casalbuoni, M. Ciminale, N. Ippolito, R. Gatto,
M. Mannarelli, and M. Ruggieri,
Rev. Mod. Phys. {\bf 86}, 509 (2014).



\bibitem{buballarev}
M. Buballa and S. Carignano, Prog. Part. Nucl. Phys. {\bf 81}, 39 (2015).




\bibitem{fullgn}
O. Schnetz, M. Thies, and K. Urlichs,
Annals Phys. {\bf 321}, 2604 (2006)

 	
\bibitem{full2}
M. Thies, J. Phys. A {\bf 39}, 12707 (2006).
 	
\bibitem{chiral11}
C. Boehmer, M. Thies, and K. Urlichs,
Phys. Rev. D {\bf 75}, 105017 (2007).

\bibitem{1d1} 	
G. Basar, G. V. Dunne, and M. Thies,
Phys. Rev. D {\bf 79}, 105012 (2009).

\bibitem{1d0}	
G. Basar and G. V. Dunne,
Phys. Rev. Lett. {\bf 100}, 200404 (2008);
Phys. Rev. D {\bf 78}, 065022 (2008).




\bibitem{symren1}	
D. Ebert, N. V. Gubina, K. G. Klimenko, 
S. G. Kurbanov, and V. Ch. Zhukovsky,
Phys. Rev. D {\bf 84},  025004 (2011).


\bibitem{symren2} 	
V. Ch. Zhukovsky, K. G. Klimenko, and I. E. Frolov, 
Moscow  Univ. Phys. Bull. {\bf 65},  539 (2010).
\bibitem{pionin}
N. V. Gubina, K. G. Klimenko, S.G. Kurbanov, and V.Ch. Zhukovsky
Phys.Rev. D {\bf 86}, 085011 (2012).

\bibitem{thieslast}
M. Thies, e-Print: arXiv:1603.06218.


\bibitem{klev}
S. P. Klevansky, Rev. Mod. Phys. {\bf 64},  649 (1992).
\bibitem{scavenius}
O. Scavenius, A. Mocsy, I. N. Mishustin, D. H. 
Rischke,
Phys. Rev. {\bf C} 64, 045202 (2001).

\bibitem{vac} 	
V. Skokov, B. Friman, E. Nakano, K. Redlich, and 
B.-J. Schaefer, 
Phys. Rev. D {\bf 82}, 034029 (2010).




\bibitem{lepage}
G. P. Lepage, http://arxiv.org/abs/hep-ph/0506330


\bibitem{mermin}
N. D. Mermin and H. Wagner, Phys. Rev. Lett. {\bf 17}, 1133 (1966).

\bibitem{cole}
S. Coleman, Commun. Math. Phys. {\bf 31}, 259 (1973).

\bibitem{j+t}J. O. Andersen and T. Brauner, Phys. Rev. D {\bf 81} 096004 (2010).
	

\bibitem{gert}
G. Aarts, D. Ahrensmeier, R. Baier, J. Berges, and J. Serreau, 
Phys. Rev. D {\bf 66}, 045008 (2002).

\bibitem{jens}
J. O. Andersen, Phys. Rev. D {\bf 75}, 065011 (2007).

\bibitem{dautry}
F. Dautry and E. M. Nyman, Nucl. Phys. A {\bf 319}, 323 (1979).






\bibitem{kut} W. Broniowski and M. Kutschera, 
Phys. Lett. {\bf B} 234, 449 (1990);
Ibid  242, 133 (1990).








 	





\bibitem{jensny}P. Adhikari and J. O. Andersen, 
e-Print: arXiv:1610.01647 [hep-th].

\bibitem{chpt}
J. Gasser and H. Leutwyler,
Annals Phys. {\bf 158}, 142 (1984).


\bibitem{eric}
E. Braaten and A. Nieto 
Eur. Phys. Journal B {\bf 11}, 143 (1999).

\bibitem{jensbose}J. O. Andersen,
Rev. Mod. Phys. {\bf 76}, 599 (2004).

 	



\end{thebibliography}
\bibliographystyle{apsrmp4-1}
\end{document}